# Imaging power of multi-fibered nulling telescopes for extra-solar planet characterization

François Hénault
UMR 6525 H. Fizeau, Université de Nice-Sophia Antipolis
Centre National de la Recherche Scientifique, Observatoire de la Côte d'Azur
Parc Valrose, 06108 Nice – France

## ABSTRACT

In this paper are discussed the nulling and imaging properties of monolithic pupil telescopes equipped with a focal plane waveguide array, which could be envisaged as precursor space missions for future nulling interferometers or coronagraphs searching for habitable planets outside of our solar system. Three different concepts of nulling telescopes are reviewed, namely the Super-Resolving Telescope (SRT) having multiple, non-overlapping exit sub-apertures and the Sheared-Pupil Telescope (SPT), either unmasked or masked with a Lyot stop placed at its exit pupil plane. For each case simple theoretical relationships allowing to estimate the nulling rate, Signal-to-Noise Ratio (SNR) and Inner Working Angle (IWA) of the telescope are established or recalled, and numerical simulations are conducted. The preliminary results of this study show that the most promising designs should either be a SRT of high radiometric efficiency associated with an adequate leakage calibration procedure, or a masked SPT with potentially deeper nulling rates but lower SNR, depending on what kind of performance is to be preferred.

## 1 INTRODUCTION

Nulling interferometers and coronagraphs are two of the most popular techniques envisaged today for the direct characterization of Extra-Solar Planets (ESP), and particularly those similar to the Earth. The first method consists in observing and spectrally analyzing the luminous power emitted by the ESP in the thermal infrared range: signs of extra-terrestrial life could then be distinguished from the measured absorption bands in the ESP atmosphere. Practically, the technique implies the realization of a space interferometer composed of several free-flying collecting telescopes, orbiting around a central combining vessel, and whose Optical Path Differences (OPD) between different interferometer arms must be equalized at the nanometer level. Additionally, the interferometer is equipped with a specific optical device, namely an Achromatic Phase Shifter (APS) turning the natural bright fringe at the centre of the interferometer Field of View (FoV) into a black, destructive fringe attenuating (or "nulling") most of the central starlight [1]. Encouraged by the recent discoveries of hundreds of exo-planets (most of them being not of terrestrial type), the European Space Agency (ESA) and National Aeronautics and Space Administration (NASA) conducted several feasibility studies for two projects of nulling interferometers, respectively named Darwin [2] and Terrestrial Planet Finder Interferometer (TPF-I [3]), leading to the conclusion that major technological improvements are necessary before launching, particularly in the domain of free-flying spacecrafts control within the required accuracy. Therefore these both extremely ambitious projects have been postponed beyond year 2020.

The other envisaged technique is coronagraphy, which only requires the use of one single, monolithic telescope. These instruments generally work in the visible or near infrared spectral range and collect the stellar radiation reflected by the ESP, which may be $10^{-10}$ times fainter than the rays directly emitted from the star in the case of an Earth-like planet. Hence the advantage of operating only one telescope in space is somewhat counterbalanced be the extremely high-required dynamic range, itself imposing very tight image quality, stray light and thermal stability control requirements. A comprehensive review of the main optical devices envisioned to block the starlight into a coronagraph system and of their compared performance can be found in Ref. [4]. It has to be noticed, however, that up to now only a few studies were devoted to the intermediate case of a single-pupil nulling coronagraph operating in the near or mid-infrared range

[5-9]: such a space borne observatory of typically 5-m diameter should be highly valuable for increasing our current knowledge of exo-planetary science, via the observation and characterization of hot Jupiter-like ESPs, and of the proto-planetary disks or exo-zodiacal dust clouds surrounding the most interesting candidate extra-solar systems. Additionally, such missions should be doubly beneficial for future space projects like Darwin or TPF-I, since they allow validating most of the critical technologies (at the noticeable exception of multiple free-flying spacecraft control) as well as the envisaged data reduction and pseudo-image reconstruction processes.

The purpose of this paper is then to examine the imaging properties of monolithic pupil, nulling telescopes equipped with a focal plane waveguide array, which could be envisaged as precursor space missions for Darwin or TPF-I projects. A general description of the instrument is first provided in section 2, where three different concepts of nulling telescopes are reviewed: they are the Super-Resolving Telescope (SRT) having multiple, non-overlapping exit sub-apertures (section 2.1) and the Sheared-Pupil Telescope (SPT) that could either be of the masked or unmasked type, depending if its exit pupil is limited by a Lyot stop or not (section 2.2). For each case, simple theoretical relationships allowing to estimate the nulling rate and Signal-to-Noise Ratio (SNR) maps are established (section 3), and their performance in terms of SNR and Inner Working Angle (IWA) are evaluated with the help of numerical simulations presented in section 4. A brief conclusion of the study is finally provided in section 5.

## 2  NULLING TELESCOPE DESIGNS

### 2.1  General description

The general principle of a monolithic pupil, nulling telescope is schematically illustrated in Figure 1. It is essentially based on the design of a Super-Resolving Telescope (SRT) originally presented in Ref. [5], to which a couple of practical improvements have been added:

1) The major change probably consists in a new focal plane arrangement, where the usually central, single mode optical fiber (SMF) is replaced with an array of single mode waveguides (SMW) allowing for multiplex measurement of the observed extra-solar planets (alternatively the central SMF could be laterally moved in the telescope focal plane by means of a mechanism). The potential benefits of this modified configuration are addressed in the following sections.

2) Moreover, all "densification" optical components formerly located before the final, multi-axial combining optics (in the **P'** plane of Figure 1) have been removed for the sake of simplicity.

Some other technical characteristics of this design are briefly summarized below:

- The collecting telescope is of Cassegrain type. Its own entrance pupil is assumed to be located on the primary mirror (plane **P** on Figure 1) and optically conjugated with the exit combining optics in plane **P'** by some relaying optics, including at least one diverging element near the telescope focal plane (see Figure 1). The beam is further re-collimated in the direction of the recombining optics.

- A servo-controlled FoV rotator, either reflective or dioptric, allows turning the destructive fringe pattern created by the nulling coronagraph around its line of sight, in order to modulate the signal emitted by an off-axis ESP. This FoV rotating subsystem is fed with error signals generated by downstream tip-tilt and wavefront sensors.

- Here we assume the Achromatic Phase Shifter (APS) to be made of dispersive glass plates [10]. Alternative designs [11-12] could also be employed with the drawback of more complicated opto-mechanical design.

- Care has been taken that the general arrangement of beamsplitters and folding mirrors is fully symmetric, ensuring that each separated beam experiences the same number of reflections and transmissions. This is in agreement with one of the most basic rule of thumb of nulling interferometry [11].

- Finally, all optical components of the exit combining optics are of modest size and could be integrated into a small structure, hence relaxing the stringent mechanical and thermal stability constraints usually applicable to space-borne interferometer arrays.

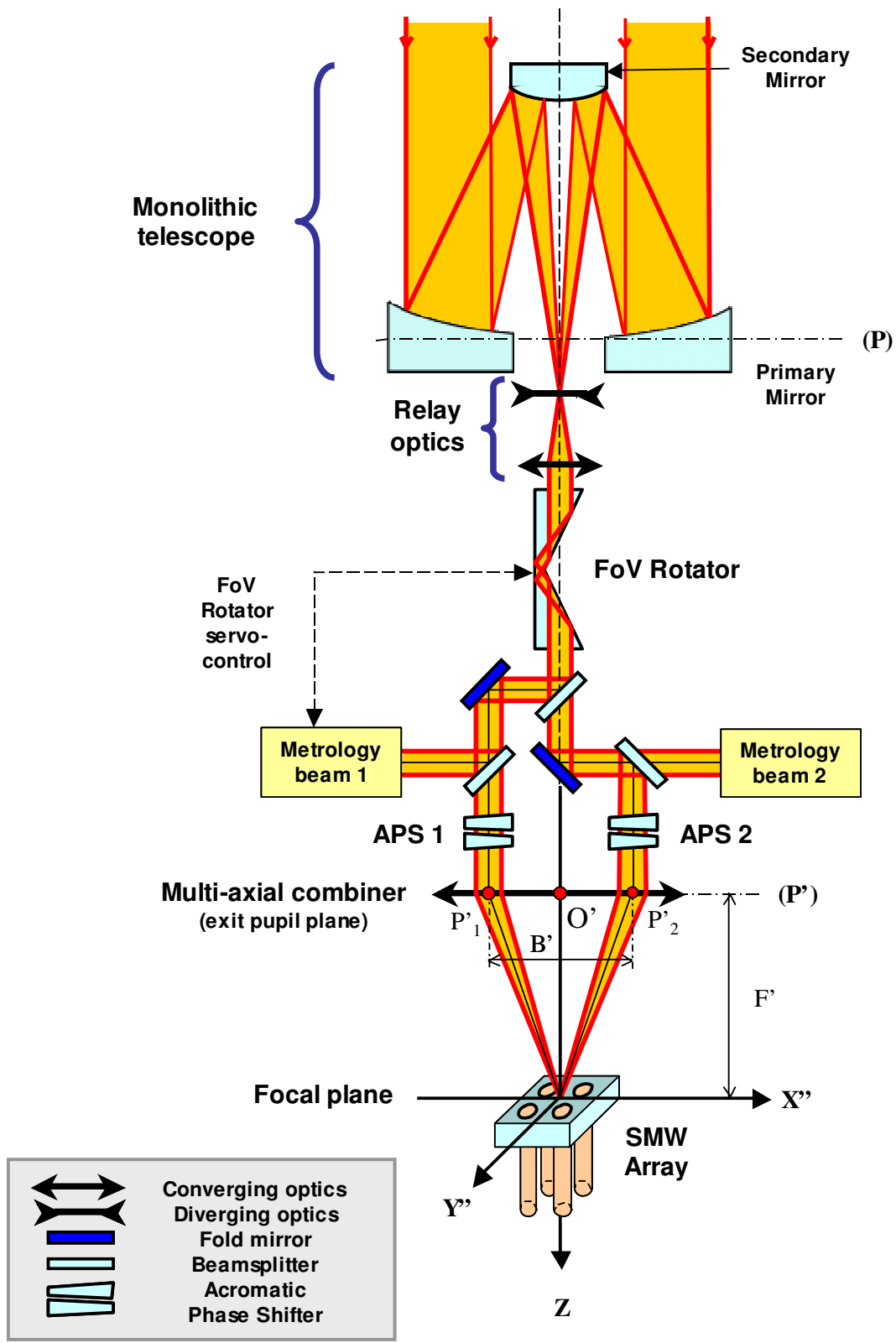

Figure 1: General design of a multi-fibered nulling telescope.

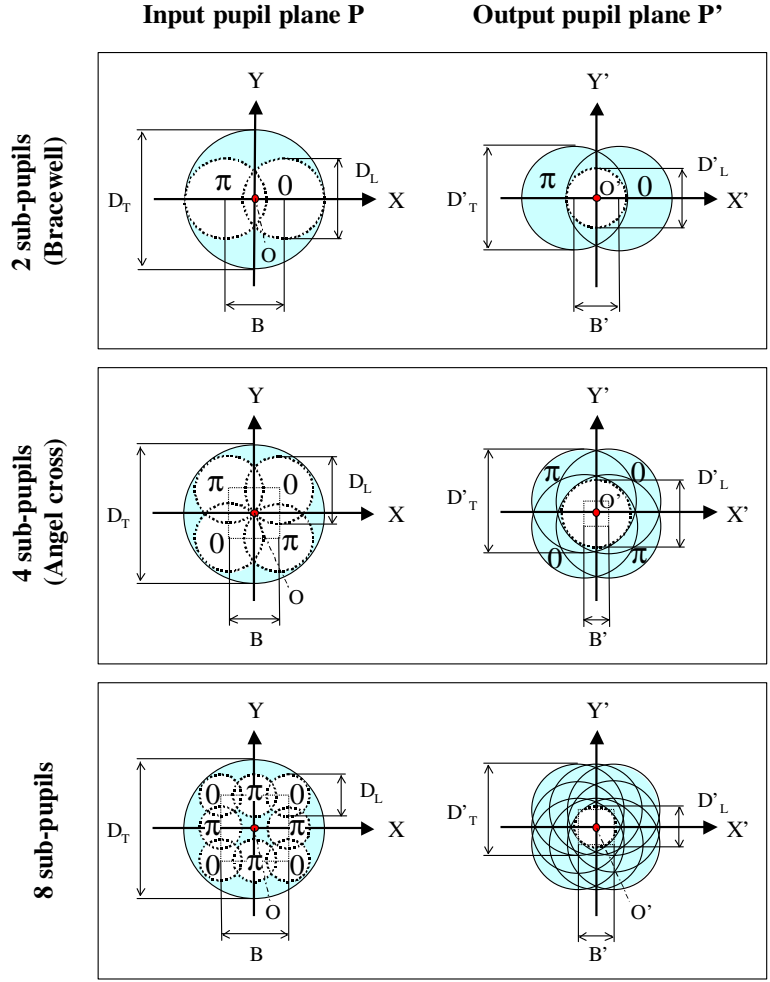

Figure 2: Sub-pupil and phase-shift arrangement for N = 2, 4 and 8 cases.

In this paper will be reviewed three different multi-aperture configurations denoted by the number N of their sub-pupils, with N = 2 (similar to Bracewell's original scheme [1] as sketched in Figure 1 and the upper row of Figure 2), N = 4 (also known as Angel cross [13], middle row of Figure 2) and N = 8 (bottom row of Figure 2). Extending the basic N = 2 configuration to N = 8 is achieved by multiplying the set of three symmetrical beamsplitters shown in Figure 1 by a factor N/2. For higher N numbers the whole opto-mechanical system would probably become unpractical, so we decided not to exceed N = 8 in this study. The geometrical parameters and phase-shift arrangements of the N = 2, 4 and 8 configurations are all defined in Figure 2.

## 2.2 Sheared-pupil telescope, with or without Lyot stop

Sheared-pupil telescopes (SPT) are an interesting alternative for nulling telescopes that has been proposed by different authors since a dozen years [6-9], and is schematized in Figure 3. With respect to the SRT, their major characteristic is that the exit sub-apertures are not fully separated in the **P'** plane, but may partly overlap as represented on the right side of Figure 2 where they are shown by blue disks. Practically, they involve interferometric setups as shown in Figure 3 (representing a Mach-Zehnder interferometer, but a Michelson equipped with cube-corners can be used as well), whose total number of needed beamsplitters should strictly be the same as for a SRT. In fact, the main distinction between both types of nulling telescopes resides in the possibility to add a Lyot stop in the exit pupil plane **P'** of the interferometer. The actual presence of this stop (indicated by white circles of diameter $D'_L$ in Figure 2) has radical impact on the imaging power of both nulling telescope families [14]:

1) With no Lyot stop laying in the **P'** plane, an "unmasked SPT" is simply equivalent to the super-resolving telescope already described in section 2.1, whose output sub-pupils may now overlap. Consequently their nulling properties shall be derived from the same theoretical relationships that will be presented in section 3.2.

2) Conversely, the presence of the Lyot stop transforms the SRT into an Axially Combined Interferometer (ACI) whose imaging properties were already discussed in Refs. [14-15]: the basic effect is to select N reduced, shifted sub-areas of diameter $D_L$ on the entrance pupil of the monolithic telescope (see Figure 2). This setup is also comparable to many past or existing multi-telescope interferometers where the beams are axially combined by means of a series of beamsplitters [16-17].

In the remainder of the text will only be distinguished the two cases above, with abbreviations "SRT" referring to both the super-resolving and unmasked sheared-pupil telescopes, and "SPT" designating a sheared-pupil telescope whose exit pupil is delimited by a Lyot stop.

## 3 THEORY

The theory of nulling telescopes has been partly exposed in Refs. [14-15]. Herein it is generalized to the case of a multi-fibered telescope where the focal plane of the instrument is equipped with SMW arrays. The employed coordinate systems and formalism are essentially the same as in Ref. [14], but the latter is summarized in section 3.1 for the sake of completeness. It must be emphasized that this formalism will only be valid if the following conditions are respected:

1) All the collecting sub-apertures in plane **P** are assumed to share identical diameters D (respectively denoted $D_T$ and $D_L$ for the SRT and SPT cases, see Figure 2), and their optical conjugates in plane **P'** also have the same diameters (noted $D'_T$ and $D'_L$).

2) This is a first-order theory: all geometrical aberrations of the collecting telescope are assumed to be negligible, as well as pupil aberrations between planes **P** and **P'**.

3) Diffraction effects are only modeled in the frame of Fraunhofer scalar diffraction theory.

### 3.1 Basic relationships

Let us consider an extended sky object subtended by a solid angle $\Omega_O$, whose angular brightness distribution is noted $O(\mathbf{s_O})$ and $\mathbf{s_O}$ is a unit vector directed at any point of the object (bold characters denoting vectors in the whole paper). Then a SMW located off-axis in the instrument focal plane should correspond on-sky to a unit vector denoted $\mathbf{s_G}$, and will transmit an optical power is proportional to:

$$T(\mathbf{s_G}) = \iint_{\mathbf{s_O} \in \Omega_O} O(\mathbf{s_O}) |\rho(\mathbf{s_O},\mathbf{s_G})|^2 \, d\Omega_O \bigg/ \iint_{\mathbf{s_O} \in \Omega_O} |G(\mathbf{s_O})|^2 \, d\Omega_O , \qquad (1)$$

where $G(\mathbf{s_O})$ is the modal function of the SMW being projected back onto the sky, and $\rho(\mathbf{s_O},\mathbf{s_G})$ is the amplitude overlap integral expressed as:

$$\rho(\mathbf{s_O},\mathbf{s_G}) = \iint_{\mathbf{s} \in \Omega_O} G^*(\mathbf{s}-\mathbf{s_G}) \, \hat{B}_D(\mathbf{s}-\mathbf{s_O}) \sum_{n=1}^{N} a_n \exp[i\varphi_n] \exp[ik(\mathbf{s_O} \, \mathbf{OP_n} - \mathbf{s} \, \mathbf{O'P'_n}/m)] \, d\Omega_O . \qquad (2)$$

The asterisk denotes the complex conjugate and other scientific notations stand for the following quantities:

$\hat{B}_D(\mathbf{s})$     The complex amplitude created at the focal plane of an individual sub-aperture of diameter $D^1$, and back-projected onto the sky. Here it is equal to $2J_1(\rho)/\rho$, where $\rho = k\,D\,\|\mathbf{s}\|/2$ and $J_1$ is the type-J Bessel function at the first order

k     The wave number $2\pi/\lambda$ of the electro-magnetic field (assumed to be monochromatic), and $\lambda$ is its wavelength

$a_n$     The amplitude transmission factor of the $n^{th}$ sub-pupil arm ($1 \leq n \leq N$)

---

[1] In all following equations, this letter stands for either $D_T$ or $D_L$, depending on what type of nulling telescope is being considered.

| | |
|---|---|
| $\varphi_n$ | The phase-shift introduced along the n$^{th}$ sub-pupil arm ($1 \leq n \leq N$) |
| $\mathbf{OP_n}$ | A vector defining the center $P_n$ of the n$^{th}$ sub-pupil in the entrance pupil plane **P** ($1 \leq n \leq N$) |
| $\mathbf{O'P'_n}$ | Correspondingly, a vector defining the center $P'_n$ of the n$^{th}$ sub-pupil in the exit pupil plane **P'** ($1 \leq n \leq N$) |
| $m$ | The optical compression factor of the system, equal to $D'_T/D_T$ for the SRT, and to $D'_L/D_L$ in the case of a SRT masked by a Lyot stop |

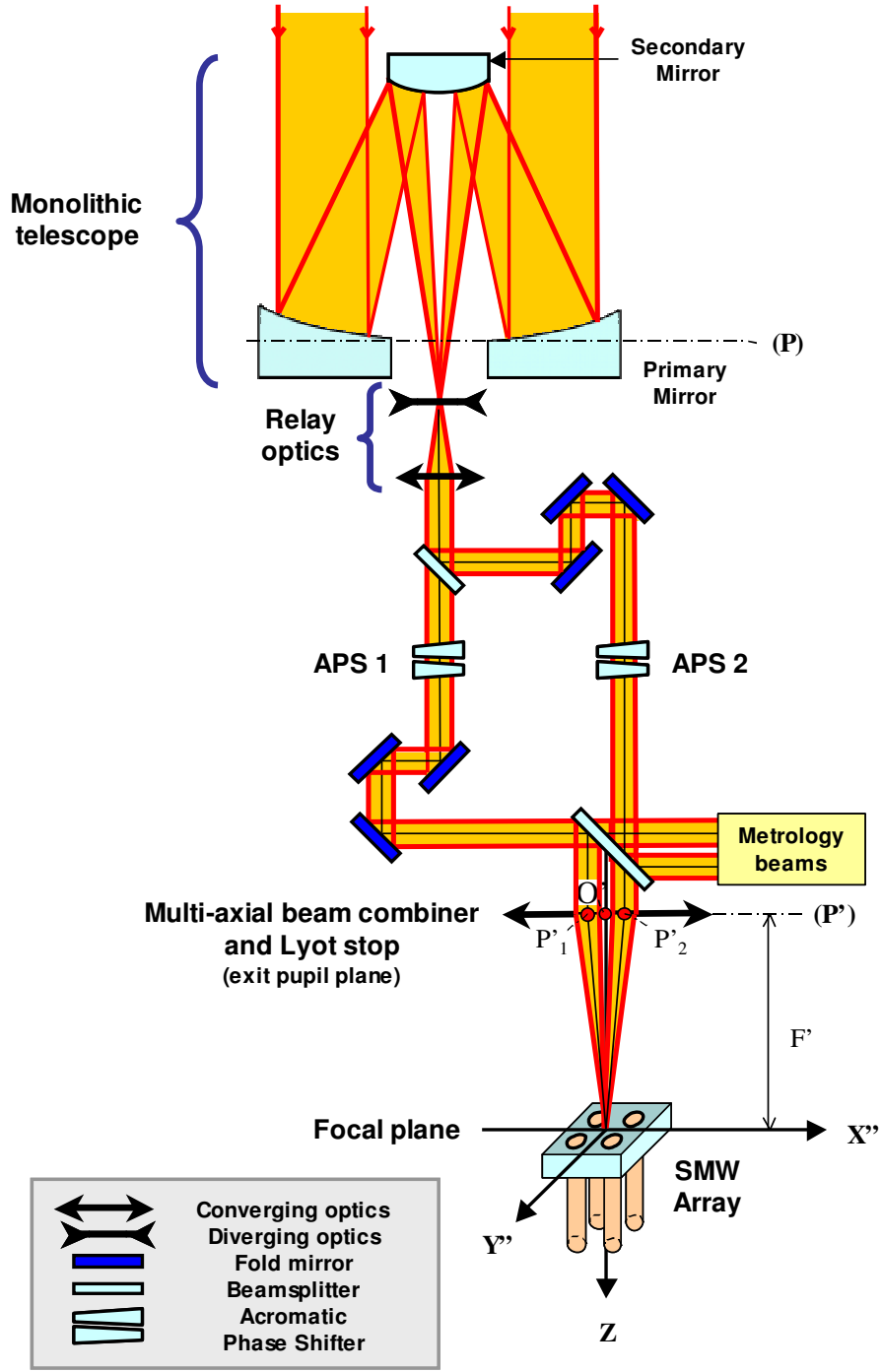

Figure 3: Sheared-pupil telescope combination scheme (FoV rotator not shown for the sake of simplicity).

This study is voluntarily restricted to the case of a very simple observed astronomical scene, being composed of a central star and an off-axis planet directed along the unit vector **s_P**, i.e.:

$$O(\mathbf{s_O}) = I_S \, \delta(\mathbf{s_O}) + I_P \, \delta(\mathbf{s_O} - \mathbf{s_P}), \tag{3}$$

where $\delta(\mathbf{s})$ is the Dirac impulse distribution, and $I_S$ and $I_P$ respectively are the intensities of the observed star and planet. Combining Eqs. (1-3) together leads to the following expressions of the total power $P_S(\mathbf{s_G})$ and $P_P(\mathbf{s_P},\mathbf{s_G})$ respectively coupled from the star and planet into a SMW pointed on-sky along vector **s_G**:

$$P_S(\mathbf{s_G}) = I_S \left| \iint_{\mathbf{s} \in \Omega_O} G^*(\mathbf{s} - \mathbf{s_G}) \, \hat{B}_D(\mathbf{s}) \sum_{n=1}^{N} a_n \exp[i\varphi_n] \exp[-i k \mathbf{s} \mathbf{O'P'_n} / m] d\Omega_O \right|^2 , \text{ and:} \tag{4}$$

$$P_P(\mathbf{s_P},\mathbf{s_G}) = I_P \left| \iint_{\mathbf{s} \in \Omega_O} G^*(\mathbf{s} - \mathbf{s_G}) \, \hat{B}_D(\mathbf{s} - \mathbf{s_P}) \sum_{n=1}^{N} a_n \exp[i\varphi_n] \exp[i k (\mathbf{s_P} \mathbf{OP_n} - \mathbf{s} \mathbf{O'P'_n} / m)] d\Omega_O \right|^2 \tag{5}$$

where the normalizing integrals of the single mode guiding function G(**s**) in Eq. (1) are omitted in order to alleviate the formulas. Although the previous theoretical relationships still look rather complicated, they can be further simplified for the both SRT and SPT nulling telescopes cases, as explained in sections 3.2 and 3.3.

## 3.2 SRT nulling maps

We first recognize in Eq. (4) the square modulus of a cross-correlation product (denoted by the symbol ⊗ in the whole paper) that should be computed very rapidly using double fast Fourier transform (FFT) algorithms:

$$P_S(\mathbf{s}) = I_S \left| G^*(\mathbf{s}) \otimes \left\{ \hat{B}_D(\mathbf{s}) \sum_{n=1}^{N} a_n \exp[i\varphi_n] \exp[-i k \mathbf{s} \mathbf{O'P'_n} / m] \right\} \right|^2. \tag{6}$$

To obtain the expression of the total power $P_P(\mathbf{s_P},\mathbf{s_G})$ coupled from the planet into the decentred optical waveguide, we should set **OP_n** = **0** in Eq. (5) for all sub-apertures with $1 \leq n \leq N$, since in that case the full telescope pupil is always utilized. This leads to the following expression of $P_P(\mathbf{s_P},\mathbf{s_G})$:

$$P_P(\mathbf{s_P},\mathbf{s_G}) = I_P \left| \iint_{\mathbf{s} \in \Omega_O} G^*(\mathbf{s} - \mathbf{s_G}) \, \hat{B}_D(\mathbf{s} - \mathbf{s_P}) \sum_{n=1}^{N} a_n \exp[i\varphi_n] \exp[-i k \mathbf{s} \mathbf{O'P'_n} / m] d\Omega_O \right|^2. \tag{7}$$

Eq. (7) basically is a Wigner distribution whose precise numerical evaluation may be quite cumbersome from a practical point of view. But assuming that **s_P** = **s_G** allows considerable simplification of the formula, now being also reducible to a cross-correlation product:

$$P_P(\mathbf{s}) \approx I_P \left| G^* \hat{B}_D(\mathbf{s}) \otimes \left\{ \sum_{n=1}^{N} a_n \exp[i\varphi_n] \exp[-i k \mathbf{s} \mathbf{O'P'_n} / m] \right\} \right|^2. \tag{8}$$

The hypothesis of equating vectors **s_G** and **s_P** is not just a mathematical artifact, because one may reasonably expect the product $G^*(\mathbf{s} - \mathbf{s_G}) \hat{B}_D(\mathbf{s} - \mathbf{s_P})$ to be negligible when both vectors **s_G** and **s_P** are angularly separated by more than one SMW core diameter. Furthermore, it corresponds to one of the most probable operating mode of this kind of nulling telescope, where the angular direction **s_P** of the extra-solar planet would have previously been determined from other methods (e.g. radial velocity or astrometry techniques), and SMW vector **s_G** consequently pointed at that precise direction. A rigorous demonstration of this assumption is however beyond the scope of the present communication.

## 3.3 SPT nulling maps

Defining the coupled power into a SMW from the star and planet is here much simpler than in the previous case. First the expression of $P_S(\mathbf{s})$ can be derived from Eq. (6), where all vectors **O'P'_n** are set to **0** ($1 \leq n \leq N$) because of the presence of the Lyot stop:

$$P_S(\mathbf{s}) = I_S \left| \sum_{n=1}^{N} a_n \exp[i\varphi_n] \right|^2 \left| G^*(\mathbf{s}) \otimes \hat{B}_D(\mathbf{s}) \right|^2. \tag{9}$$

Practically, this means that the starlight is nulled by a factor of $N_S = \left| \sum_{n=1}^{N} a_n \exp[i\varphi_n] \right|^2$ inside an angular envelope defined by function $\left| \hat{B}_D \otimes G^*(\mathbf{s}) \right|^2$. But since the amplitude and phase parameters $a_n$ and $\varphi_n$ are usually optimized to get $N_S = 0$, one should expect $P_S(\mathbf{s})$ to be uniformly null in the whole FoV. Secondly, the expression of the planet power $P_P(\mathbf{s})$ coupled into the SMW is directly derived from Eq. (5), where $\mathbf{O'P'_n} = \mathbf{0}$ for $1 \leq n \leq N$, and the same approximation than in section 3.2 holds (i.e. the SMW is assumed to be pointed at the planet).

$$P_P(\mathbf{s}) \approx I_P \left| \sum_{n=1}^{N} a_n \exp[i\varphi_n] \exp[i k \mathbf{s} \mathbf{OP_n}] \right|^2 \left| \iint_{\mathbf{s} \in \Omega_O} G^*(\mathbf{s}) \hat{B}_D(\mathbf{s}) d\Omega_O \right|^2, \tag{10}$$

where the second term of the product is a constant factor that can be assumed equal to 1. Here is retrieved one of the most basic formulas of the nulling map generated by an ACI and projected onto the sky [18]. It may seem at first glance that the expressions found for $P_S(\mathbf{s})$ and $P_P(\mathbf{s})$ for the SPT configuration are much more favorable, because they allow (at least theoretically) a full extinction of the central star, on the one hand, and the observable FoV is not limited by any envelope function, on the other hand. However this preliminary conclusion may be somewhat moderated when the radiometric properties of both types of nulling telescopes are taken into account, as described in the following section.

### 3.4 Radiometric efficiency and SNR

The radiometric performance of nulling interferometers or telescopes is already well covered by the scientific literature, from which one should for example refer to both papers [19] and [20]. Here we shall not follow the same detailed and meticulous approaches than these authors, but will rather group the main known sources of errors under a few generic categories, and make their amplitudes vary as a whole. The performance of a nulling telescope searching for ESPs or exo-zodiacal clouds can basically be expressed in terms of its Signal-to-Noise Ratio (SNR), herein written under the simplified form:

$$\text{SNR}(\mathbf{s}) \approx \frac{P_P(\mathbf{s}) \, \eta \, A \, \tau}{\sqrt{[P_S(\mathbf{s}) + N_0 \, I_S + I_{EZ}] \eta \, A \, \tau + \sigma_N^2}}. \tag{11}$$

It must be noted that this SNR map depends on the expected ESP direction $\mathbf{s}$, since it involves the expressions of the star and planet nulling maps $P_S(\mathbf{s})$ and $P_S(\mathbf{s})$ previously established in sections 3.2 and 3.3, themselves depending on $\mathbf{s}$. The other scientific notations in relation (11) are describing the following physical parameters:

- $\eta$     The radiometric efficiency of the whole system, including among others mirrors reflectivity, beamsplitters transmission and the quantum efficiency of the detectors
- A     The effective collective area of the nulling telescope. For the SRT it is obviously equal to $\pi D_T^2/4$, but in the case of a SPT limited with a Lyot stop, it is obtained by adding all the white areas of diameter $D_L$ shown on the left side of Figure 2. Hence A depends on the actual value of the shear B separating the input sub-pupils, further named "equivalent entrance baseline" in the paper. There is generally no simple analytical expression for A and this parameter must be integrated numerically
- $\tau$     The total integration time on the detector, which may typically attain several hours for an ESP observing mission
- $N_0$     The "null floor" of the instrument, i.e. the minimal value of $N_S$ (as defined in section 3.3) that can be achieved depending on the instrumental imperfections: the latter typically include OPD instability, amplitude and polarization mismatches or residual chromaticity inherent to the optical materials or coatings. Extensive evaluation and budgets of this crucial parameter have been described in Refs. [19-20]
- $I_S$     The luminous intensity of the central star, as mentioned in § 3.1
- $I_{EZ}$     The luminous intensity of a zodiacal dust cloud present in the observed stellar system, acting as a diffuse background and potentially hindering ESP detection. This parameter can also be used to model the thermal self-emission of the optics
- $\sigma_N$     The noise generated by the detection unit (e.g. CCD read-out noise)

The next section is devoted to the comparison of some typical performance of the SRT and SRT as function of all the above parameters, obviously in terms of SNR but also of their accessible Inner Working Angle (IWA).

# 4 NUMERICAL SIMULATIONS

## 4.1 Principle of numerical simulations

The principle of the herein presented numerical simulations is illustrated in Figure 4 and Figure 5. The key performance of a nulling telescope is assumed to be its SNR, basically evaluated from relation (11) where the intensities of the star and planet $P_S(\mathbf{s})$ and $P_P(\mathbf{s})$ are deduced from Eqs. (6)-(8) or (9-10) respectively, depending on the basic nature of the telescope (being of the SRT or SPT types). The main employed input parameters are summarized in Table 1.

Figure 4 first shows the results of a preliminary sensitivity study of the maximal SNR that can be achieved in the FoV, as function of six of the most critical parameters of Table 1, namely (from left to right and top to bottom in the Figure) the magnitude of the target star allowing to define its intensity $I_S$, the integration time on the detector $\tau$, the planet/star and background/star intensity ratios $I_P / I_S$ and $I_{EZ} / I_S$, the "null floor" $N_0$ and the detection noise $\sigma_N$. Searching for typical SNR around 3 or 4 in the case of 5-m class telescope looking for hot Jupiters and exo-zodiacal clouds in the IR thermal range, we finally select $\tau = 10$ mn, $N_0 = 10^{-4}$, $\sigma_N = 20$ e⁻ and a target star of magnitude 8 as a reasonable set of starting parameters for the next series of simulations.

| PHYSICAL PARAMETERS | VALUES |
|---|---|
| Maximal telescope diameter | $D_T = 5$ m |
| Telescope focal length | $F = 50$ m |
| Reference wavelength | $\lambda = 10$ µm |
| Optical compression factor | $m = 1/500$ |
| Total number of sub-pupils | $N = 2, 4$ and $8$ |
| Equivalent entrance baseline | $0 \leq B \leq 10$ m |
| Overall radiometric efficiency | $\eta = 0.1$ |
| Total integration time | $\tau = 10$ mn |
| Null floor | $N_0 = 10^{-4}$ |
| Irradiance from central star $I_S$ | Equivalent to magnitude = 8 in thermal IR band |
| Planet / star intensity ratio | $I_P / I_S = 10^{-4}$ |
| Background / star intensity ratio | $I_{EZ} / I_S = 10^{-3}$ |
| Detection noise | $\sigma_N = 20$ e⁻ |

Table 1: Main physical parameters used in numerical simulations.

Figure 5 illustrates the principal computation steps being carried out when simulating a single SNR map, for both the nulling SRT (Figure 5a-g) and SPT configurations (Figure 5h-l). Figure 5a shows the exit pupil map of the SRT after having been transferred back into the entrance pupil plane $\mathbf{P}$ (i.e. enlarged by a factor $1/m$). Figure 5b-c display grey-scale and three-dimensional (3D) views of the nulling map $P_P(\mathbf{s})$ transmitted by the SRT, evaluated from Eq. (8) – note the dotted grey arrow indicating the IWA of the instrument. Figure 5d-e show the star leakage map $P_S(\mathbf{s})$ generated by the SRT in the whole FoV, computed from Eq. (6). Finally, Figure 5f-g present the SNR map achieved by the SRT, showing distinctly four off-axis areas in the FoV where the SNR is maximal, and consequently planet detection and characterizations are deemed the most favorable.

Figure 5h-l basically present the same illustrations than for the SRT configuration with only three minor differences:

- The program automatically determines the useful input sub-apertures on the telescope entrance pupil due to the presence of the Lyot stop (Figure 5h), and then evaluates the effective collecting area A of the SPT digitally.

- The nulling map $P_P(\mathbf{s})$ transmitted by the SPT is now directly computed from relation (10).

- The star leakage map $P_S(\mathbf{s})$ is not shown since it now reduces to the constant null floor value $N_0$.

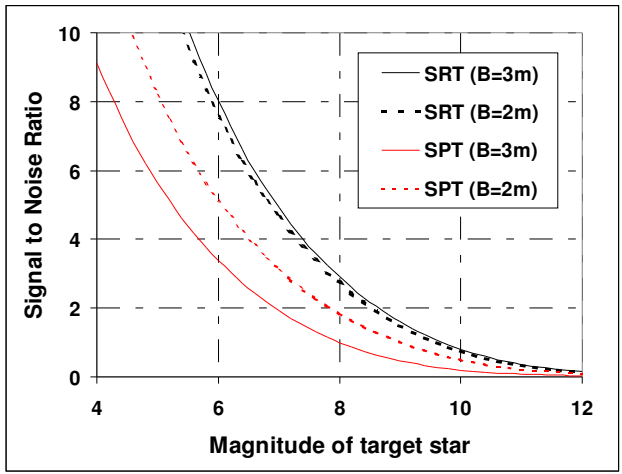
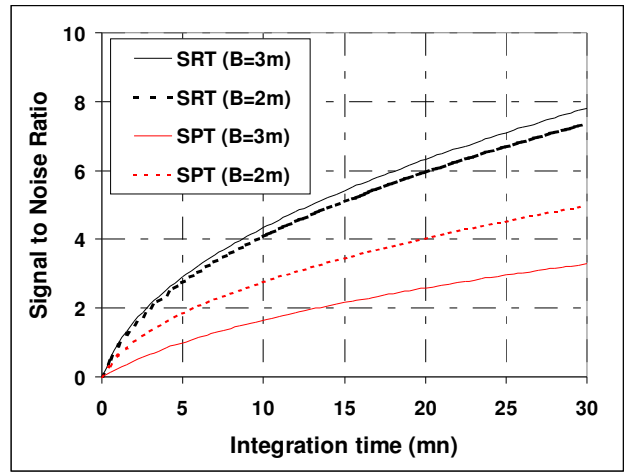
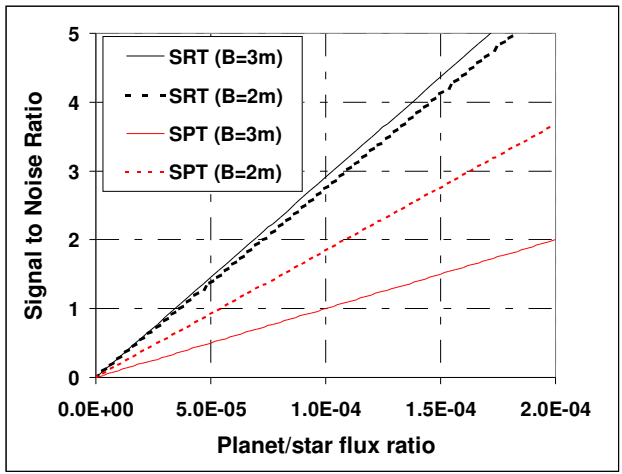
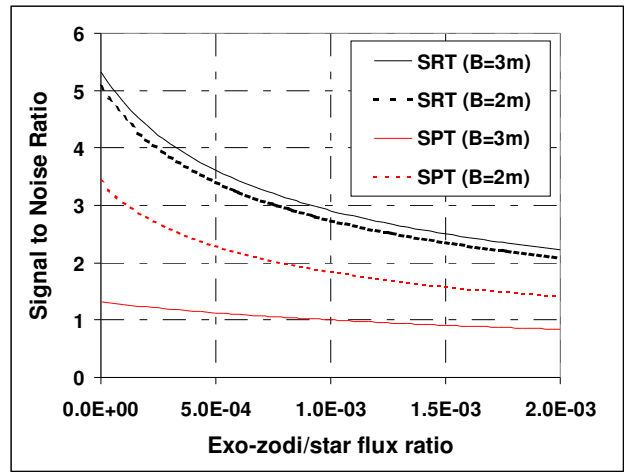
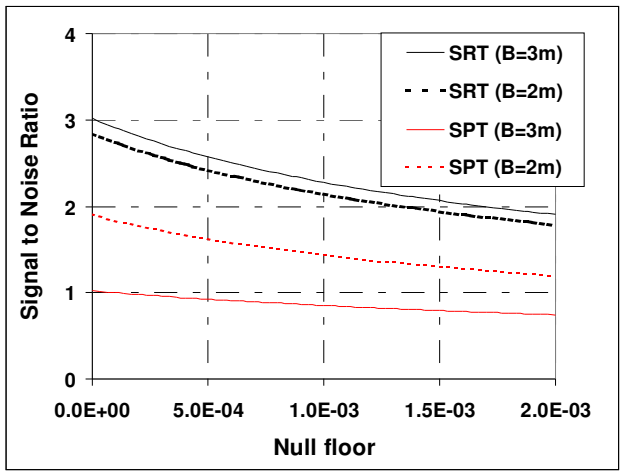
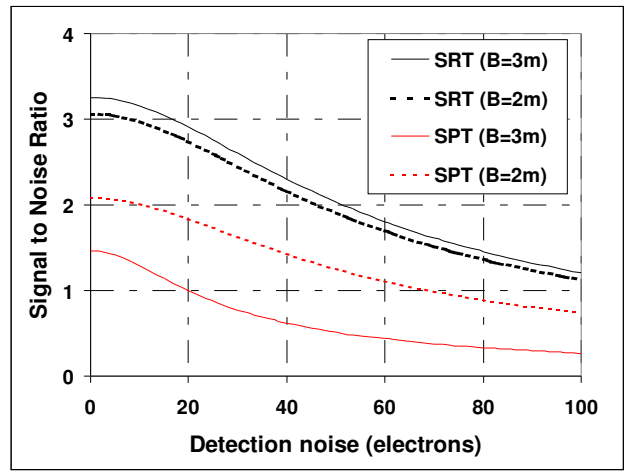

Figure 4: SNR sensitivity curves for SRT and SPT having equivalent entrance baselines B = 2 and 3 m (N=8 sub-pupils).

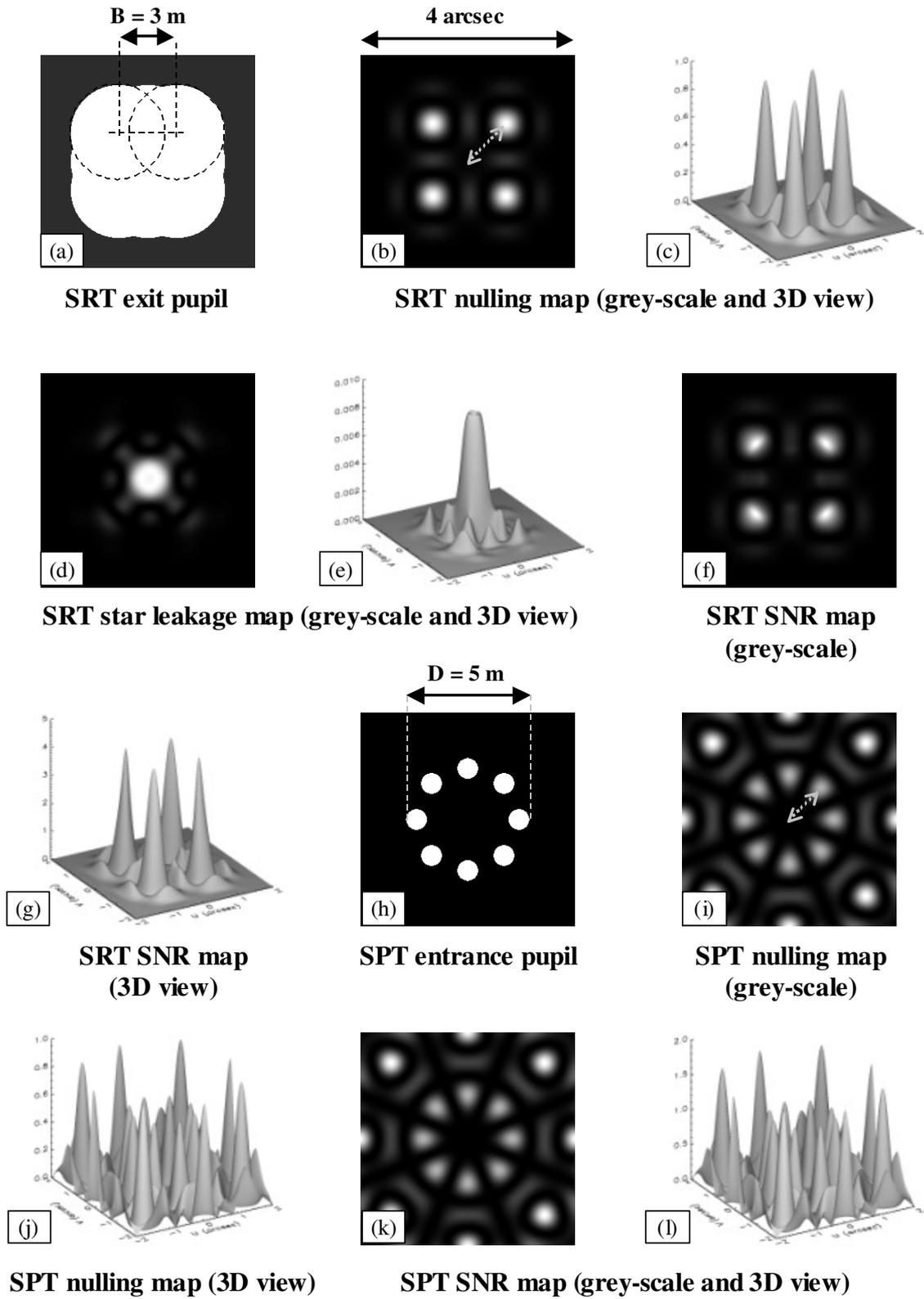

Figure 5: Illustrating SRT and SPT numerical simulations (case of 8 sub-pupils, equivalent entrance baseline B = 3 m).

It must be noticed that the equivalent entrance baseline is here equal to B = 3 m, which is a very unfavorable case for the SPT since its entrance sub-pupils are reduced significantly, but has been chosen here for illustration purposes. In the next section are now reviewed the final performance of both types of nulling telescopes, as a function of their equivalent entrance baseline B varying from 0 to 10 m[1].

### 4.2 Analysis of the results

The final results of this preliminary study are presented in Figure 6. From left to right and top to bottom, it shows the radiometric efficiency, expected SNR, Inner Working Angle (IWA) and minimal extinction rate curves, computed for the both SRT and SPT in their N = 2, 4 and 8 configurations as function of their real or equivalent entrance baseline B. These four different aspects of a nulling telescope performance were calculated as follows:

1) Curves in the upper left panel of the Figure actually correspond to a relative efficiency, being equal to the product of the SMW coupling ratio by a factor $A/(\pi D_T^2/4)$ standing for the useful fraction of the total optical power collected by the telescope (coefficient η in Table 1 is not taken into account). For each considered case the SMW coupling efficiency has been maximized digitally with respect to the exact geometry of the exit pupil.

2) SNR curves on the upper right panel are basically computed from Eq. (11) as explained in the previous sub-section.

3) IWA is classically defined as the minimal angular distance from the star where the planet throughput is higher than 50 % of its maximal value in the whole FoV [4]. It is numerically computed for each considered case as presented on the lower left panel of Figure 6 (also illustrated by dotted grey arrows in Figure 5b and Figure 5i).

4) Finally, the worst extinction ratio attained in the FoV of both types of nulling telescopes, simply defined as $P_S(\mathbf{s})/I_S + N_0$, is illustrated on the lower right panel of Figure 6.

Examining these curves reveals that both telescopes have similar performance in terms of SNR and radiometric efficiency (top panels of the Figure) for relatively small equivalent entrance baselines B, typically B ≤ 2 m. For higher values of B the smaller effective collecting area in its entrance pupil penalizes the SPT with respect to the SRT. Furthermore, the SPT is limited by a "cut-off" B number (here equal to 5 m for the Bracewell-like configuration, when the intersection between all sub-pupils becomes null), while the SRT continues to operate well beyond this limit. Here is perhaps the main drawback of the SPT arrangement since it can only work in a region where its IWA is typically higher than 0.5 arcsec (see the bottom left panel of the Figure). The possibility to access smaller IWAs is therefore the major advantage of the SRT[2]. However this benefit is counterbalanced by a non-uniform nulling ratio in the SRT FoV, being increased by stellar leaks (bottom right panel of the Figure): this may occur in significant FoV areas with the detrimental consequence that SMWs can only be set at very precise locations in the SRT focal plane as illustrated in Figure 5f-g. A possible remedy could be to implement a star leakage calibration procedure, using those SMWs that are "blinded" by the starlight as reference. In conclusion, it seems that both the multi-fibered SRT and SPT are probably not well suited for a TPF-I/Darwin-like space mission in search of Earth-like planets, but should have the capacity to hunt for hot Jupiters and exo-zodiacal clouds as demonstrated by the simulations presented above. This study also confirms that SRTs and SPTs have adverse advantages and drawbacks: to benefit of all imaging and nulling capacities offered by the both concepts, one may think of installing a Lyot stop of adjustable diameter in the exit pupil plane of the telescope, turning it into a convertible and versatile ESP observatory.

## 5    CONCLUSION

Space borne multi-fibered, nulling telescopes are excellent candidates for certain types of scientific missions, such as the characterization of hot Jupiter ESPs or exo-zodiacal dust clouds surrounding potentially habitable extra-solar systems. In this paper, we did not try to define precise technical requirements for such an instrument, but simply to compare the two main options regarding their optical design with the help of explicit theoretical relationships and numerical simulations. The results of this preliminary study confirm that both types of nulling telescopes exhibit adverse advantages and drawbacks, because the SRT can typically achieve higher radiometric efficiency and smaller IWAs but suffers from modest and non-uniform nulling rates, while the SRT has better extinction capacity but is inferior in terms of SNR and

---

[1] However, cases when B → 0 m are probably of no interest here, since the SRT and SPT would reduce to a "full pupil nulling telescope" having an infinite dark FoV as discussed in Ref. [14], § 3.3.1.
[2] Except for the 8 sub-pupils configuration: this is a probable consequence of its natural extended dark FoV.

IWA. Hence it may be concluded that the most promising designs might either be a SRT associated with an adequate leakage calibration procedure, or a Lyot-masked SPT having reduced radiometric efficiency but unsurpassable nulling power, depending on what performance is to be preferred.

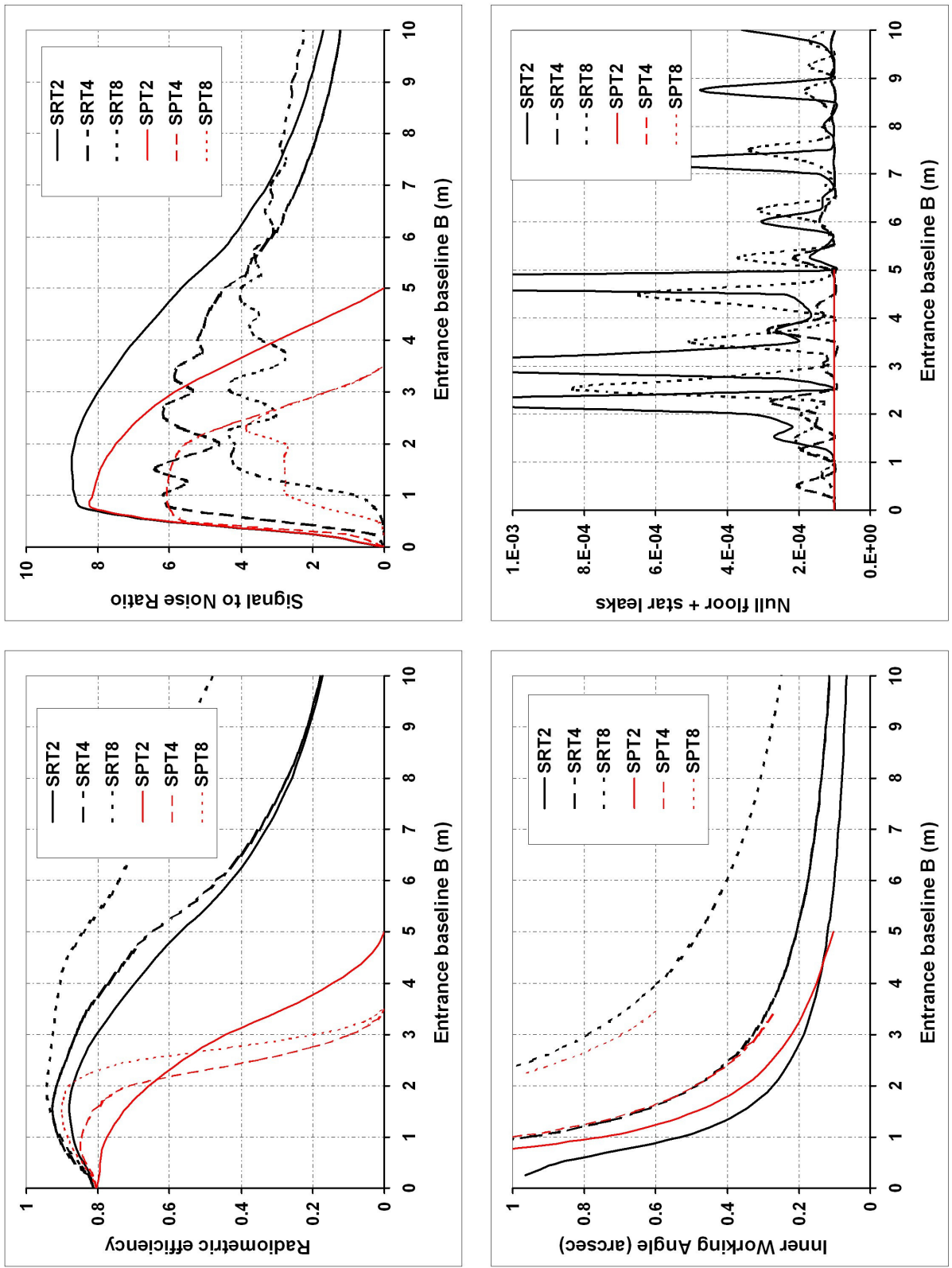

Figure 6: Compared radiometric performance of the SRT and SPT (N = 2, 4 and 8 sub-pupils).